\documentclass[twocolumn,showpacs,preprintnumbers,amsmath,amssymb]{revtex4}

\usepackage{dcolumn}
\usepackage{bm}
\usepackage{epsfig}

\usepackage{ulem}%
\usepackage{color}%

\begin{document}


\title{Possible existence of a filamentary state in type-II superconductors} 

\author{V. Kozhevnikov$^{1}$, A.-M. Valente-Feliciano$^{2}$, P. J. Curran$^{3}$, A. Suter$^{4}$, A. H. Liu$^{5}$, G. Richter$^{6}$,  E. Morenzoni$^{4}$, S. J. Bending$^{3}$ and C. Van Haesendonck$^5$}
\affiliation{
$^1$Tulsa Community College, Tulsa, Oklahoma 74119, USA\\
$^2$Thomas Jefferson National Lab, Newport News, VA 23606, USA\\
$^3$University of Bath, Bath BA2 7AY, United Kingdom\\
$^4$Paul Scherrer Institut, 5232 Villigen PSI, Switzerland\\
$^5$Solid State Physics and Magnetism Section, KU Leuven, BE-3001 Leuven, Belgium\\
$^6$Max-Planck-Institut for Intelligent Systems, 70569 Stuttgart, Germany
}.\\


\begin{abstract}
\noindent The standard interpretation of the phase diagram of type-II superconductors was developed in 1960s and has since been considered a well-established part of classical superconductivity. However, upon closer examination a number of fundamental issues arise that leads one to question this standard picture.   To address these issues we studied equilibrium properties of niobium samples near and above the upper critical field $H_{c2}$ in parallel and perpendicular magnetic fields. The samples investigated were very high quality films and single crystal discs with the Ginzburg-Landau parameters 0.8 and 1.3, respectively. A range of complementary measurements have been performed, which include dc magnetometry, electrical transport, $\mu$SR spectroscopy and scanning Hall-probe microscopy. Contrarily to the standard scenario, we observed that a superconducting phase is present in the sample bulk above $H_{c2}$ and the field $H_{c3}$ is the same in both parallel and perpendicular fields. Our findings suggest that above $H_{c2}$ the superconducting phase forms filaments parallel to the field regardless on the field orientation. Near $H_{c2}$ the filaments preserve the hexagonal structure of the preceding vortex lattice of the mixed state and the filament density continuously falls to zero at $H_{c3}$. Our work has important implications for the correct interpretation of properties of type-II superconductors and can also be essential for practical applications of these materials.

\end{abstract}\


\maketitle

Interpretation of equilibrium properties of superconductors has a pivotal significance for the entire realm of quantum physics, extending from neutron stars to the standard model \cite{Huebener, Ranninger}. Therefore it is important to verify any concern related to  description of these properties.

Type-II superconductors subjected to a magnetic field $H$ above the lower
critical field $H_{c1}$ can be found in three equilibrium states \cite{De Gennes, L&P, Abrikosov, Tinkham}: in the mixed state (MS), where average magnetic induction $\bar{B}< H$ and currents form vortices organized in a hexagonal lattice; in a ``surface superconductivity" state, where $B=H$ everywhere except a
sheath with thickness of the order of the Ginzburg-Landau (GL) coherence length  near the surface parallel to $H$; and in the normal state (NS). The typical phase diagram of  type-II superconductors of cylindrical geometry (such as, e.g.,  infinite circular cylinders and slabs with thickness greatly exceeding the penetration depth) in parallel magnetic field, or of massive samples with demagnetizing factor $\eta=0$ \cite{Landafshitz_II} is shown in Fig.~1.  Transitions between states, occurring at the critical fields $H_{c2}$ and $H_{c3}$,  are second order phase transitions. In ellipsoidal samples with
$\eta\neq0$ the sheath formes an equatorial band whose width decreases with increasing $\eta$. In samples with $\eta=1$ (infinite slabs in perpendicular field) the band vanishes and surface superconductivity disappears. Since MS in such samples starts from $H = (1-\eta)H_{c1}=0$, their phase diagram consists of a single curve $H_{c2}$.
\begin{figure}
\vspace{15 mm}
\epsfig{file=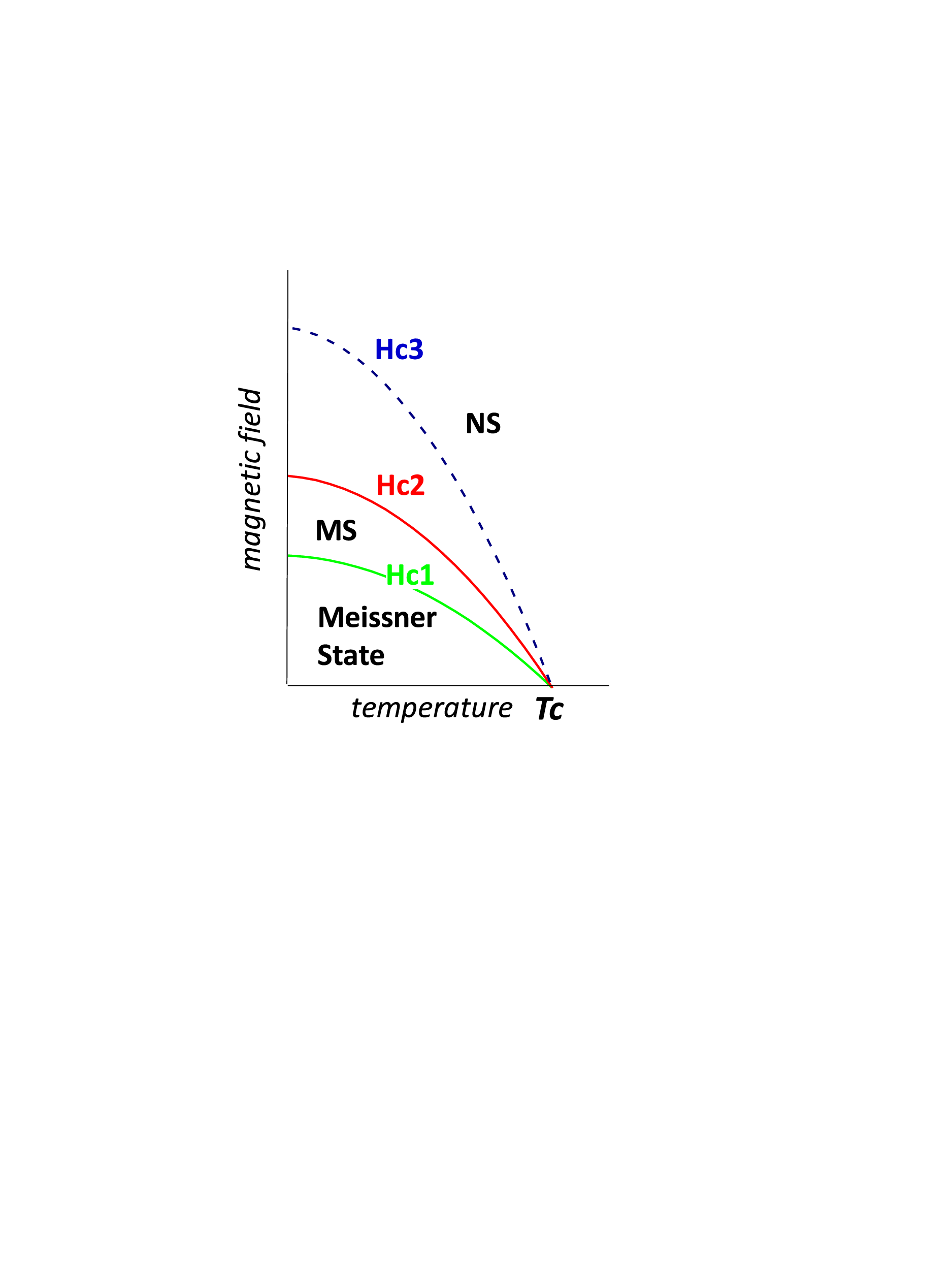,width=3.7 cm}
\caption{\label{fig:epsart} Phase diagram of a massive type-II superconductor of cylindrical geometry in parallel magnetic field. MS and NS denote the mixed and the normal states, respectively.} 
\end{figure}

This interpretation of the properties of type-II superconductors is based on two well known solutions of the linearized GL equation obtained by Abrikosov \cite{Abrikosov57} and Saint-James and de Gennes \cite{Saint-James}. In spite of a narrow range of applicability of the GL theory \cite{GL, Shoenberg, Gorkov2}, its tremendous success has been due to an explanation of very puzzling properties of these materials.

\begin{figure}
\epsfig{file=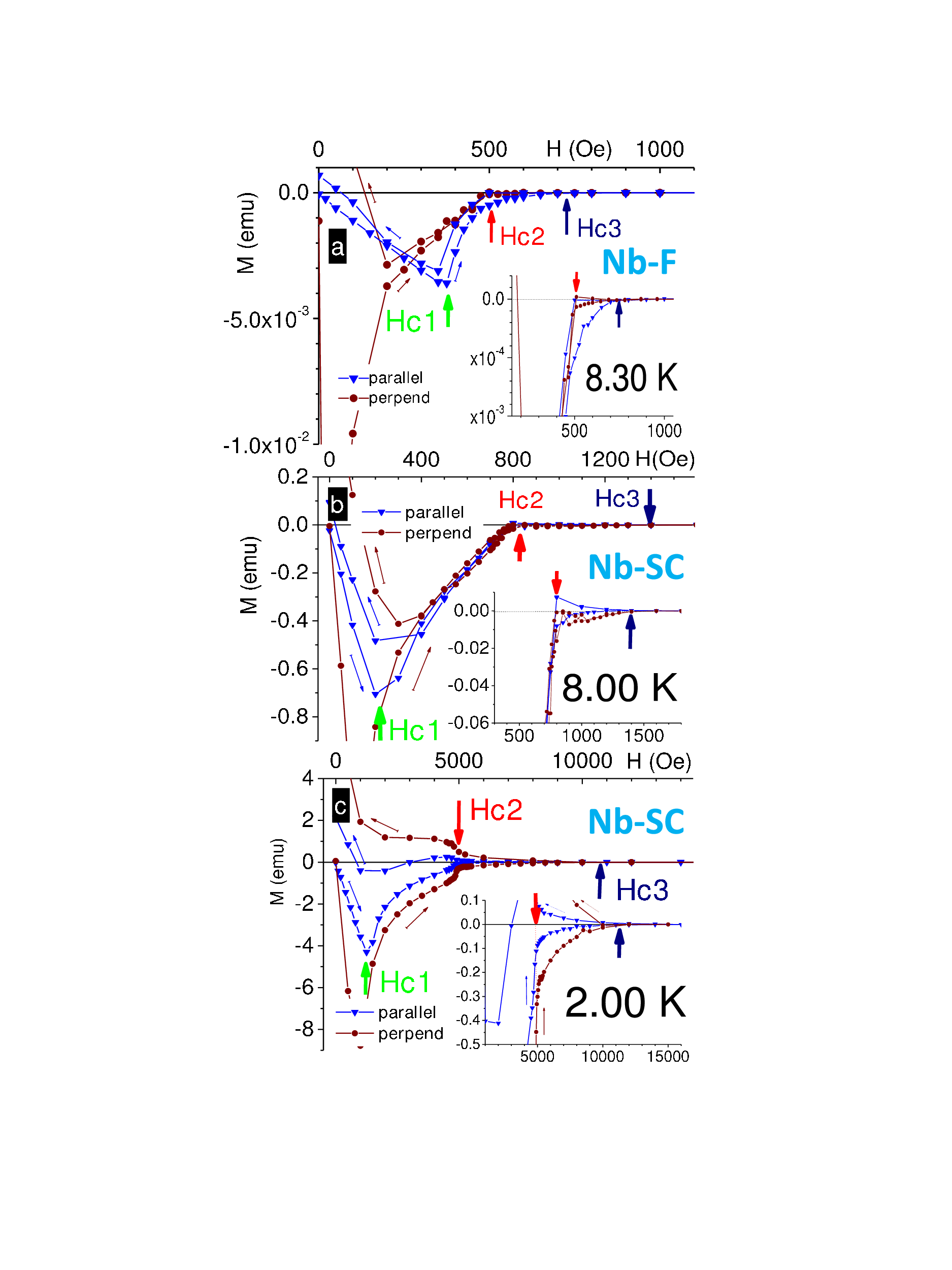,width=5.4 cm}
\caption{\label{fig:epsart} Magnetic moment of Nb-F and Nb-SC in parallel and perpendicular fields at indicated temperatures. Inserts: same data on an enlarged scale.} 
\end{figure}

One of such puzzles was a factor of two discrepancy in the upper critical field following from magnetic and resistive measurements. It was often attributed to defects  and insufficient sensitivity of magnetometers (see, e.g. \cite{Druyvesteyn}) and therefore  ignored in theories (e.g., \cite{Tinkham63}). Saint-James and de Gennes treated superconductivity above $H_{c2}$ as an equilibrium property, thus providing an interpretation of the entire phase diagram within 
one theory.

However this standard picture raises some questions. In particular, it implies that in parallel geometry superconductivity nucleates at a field ($H_{c3}$) almost twice as large as the field at which it nucleates in perpendicular geometry ($H_{c2}$). By definition, the field passes the sample in the NS being unperturbed, i.e. not noticing the surface. Hence, nucleation at $H_{c3}$ should not depend on the field-to-surface orientation. Also, in this scenario the states coexisting at $H_{c2}$ belong to different classes of symmetry, like crystal and liquid. Hence, the phase transition at $H_{c2}$ should not be of second order \cite{LL}. In particular, in samples with $\eta=1$ the coexisting states are the MS and the NS. Apart from different symmetries, the minimum amount of a superconducting (S) phase needed to create the vortex lattice is $\thickapprox$\,10\% of the sample volume. Hence, this transition should not be continuous.

In this communication we challenge the standard interpretation of the phase diagram of type-II superconductors by showing
that above $H_{c2}$ the S phase forms filaments parallel to applied field regardless of its orientation.

To address the indicated questions, we measured magnetization, electrical transport, $\mu$SR spectra and took scanning Hall-probe microscopy (SHPM) images on Nb samples. Those were two high-purity 5.7\,$\mu$m thick films 4$\times$6\,mm$^2$ (Nb-F) and 2$\times$4\,mm$^2$ (Nb-F2), and two one-side polished 1\,mm thick discs with diameter 7\,mm (Nb-SC) and 19\,mm (Nb-SC2) cut from the same single crystal rod. The film samples were cut from a film grown on sapphire using electron cyclotron resonance technique \cite{Anne-Marie}; its residual resistivity ratio is 640. The GL parameter $\kappa$ determined from magnetization curves in parallel field is 0.8 (1.3) near the critical temperature $T_c$ rising up to 1.1 (1.6) at 2 K for the Nb-F (Nb-SC) sample. $T_c$ of the film (single crystal) samples is 9.25 K (9.20 K). As verified by magnetization measurements, the samples are nearly pinning-free at $T\gtrsim$ 8 K. 

The magnetic moment $M$ was measured on the Nb-F and Nb-SC samples using Quantum Design MPMS system. Typical data for high temperatures are shown in Figs.\,2a, b. We see that $H_{c2}$ and $H_{c3}$ are well distinguishable for both samples. At low temperatures flux trapping is more significant, however it is still possible to resolve the critical fields. An example of the low-temperature data for the Nb-SC sample is shown in Fig.\,2c. We observe that above $H_{c2}$ the S phase is present for both field orientations, and in both cases $H_{c3}$ is the same. These results are inconsistent with the surface sheath interpretation. In particular, they suggest that 
above $H_{c2}$ the S phase forms either droplets or filaments with decreasing number density under increasing field.
\begin{figure}
\epsfig{file=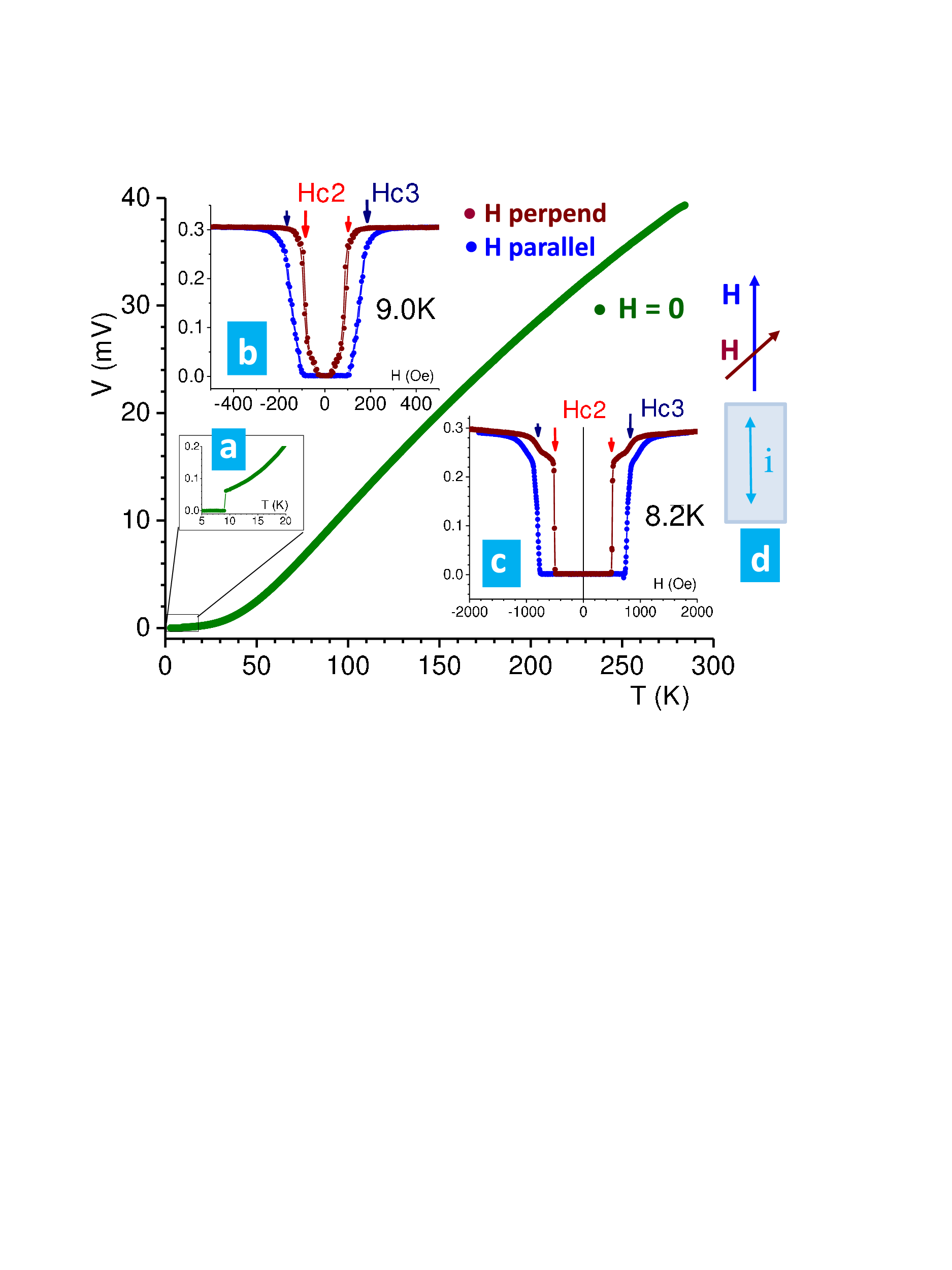,width=5.5 cm}
\caption{\label{fig:epsart} Voltage $V$ across the Nb-F2 sample. Green dots represent $V(T)$
at $H=0$; in (a) these data are shown on a magnified scale. (b) and (c): $V(H)$ obtained in
parallel (blue dots) and perpendicular (brown dots) fields at indicated temperatures; the red and navy arrows indicate $H_{c2}$ and $H_{c3}$ inferred from the $M(H)$ data, respectively. (d): current (\textit{i}) and the field configurations.  }
\end{figure}

The electrical resistance was measured for Nb-F2 sample using a low-current (2 mA)
ac bridge. Voltage across potential leads measured \textit{vs} $T$ at $H=0$ and \textit{vs} $H$ at constant $T$ is shown in Fig.\,3. In Figs.\,3\textit{b} and 3\textit{c} we see that resistance drops abruptly at $H_{c3}$ in parallel and at $H_{c2}$ in perpendicular field, where $H_{c2}$ and $H_{c3}$ are inferred from $M(H)$. This is in line with the data on electrical transport used to support the surface superconductivity interpretation (see, e.g., \cite{Hempstead,
Gygax, Cardona, Deutscher}).  However, this interpretation conflicts with $M(H)$ data. At the same time both resistance and $M(H)$ are consistent with a filament scenario, provided the filaments are parallel to the applied field. The resistance data rule out the droplet scenario.

Alternatively, magnetic properties can be studied by $\mu$SR. Its bulk version makes use of
$4 \, {\rm MeV}$ polarized muons, probing $B$ at $\sim$\,0.1 mm below the sample surface, i.e. in the bulk (see, e.g., \cite{NL, Sonier, Yaouanc} for details).

$\mu$SR spectra were acquired for the Nb-SC2 sample in perpendicular field at the Dolly instrument of the Swiss Muon Source.  Number of events of muon decays collected in each data point is $3\cdot10^6$; statistical error in measured field is $\lesssim0.1\%$. Typical time-spectra for the MS are shown in Fig.\,4a, where insert shows the spectra for the NS. For comparison, Fig.\,4b shows the spectra for the IS taken at the same reduced temperature and field for a type-I In  sample; the insert shows the spectra of In in the NS.
\begin{figure}
\epsfig{file=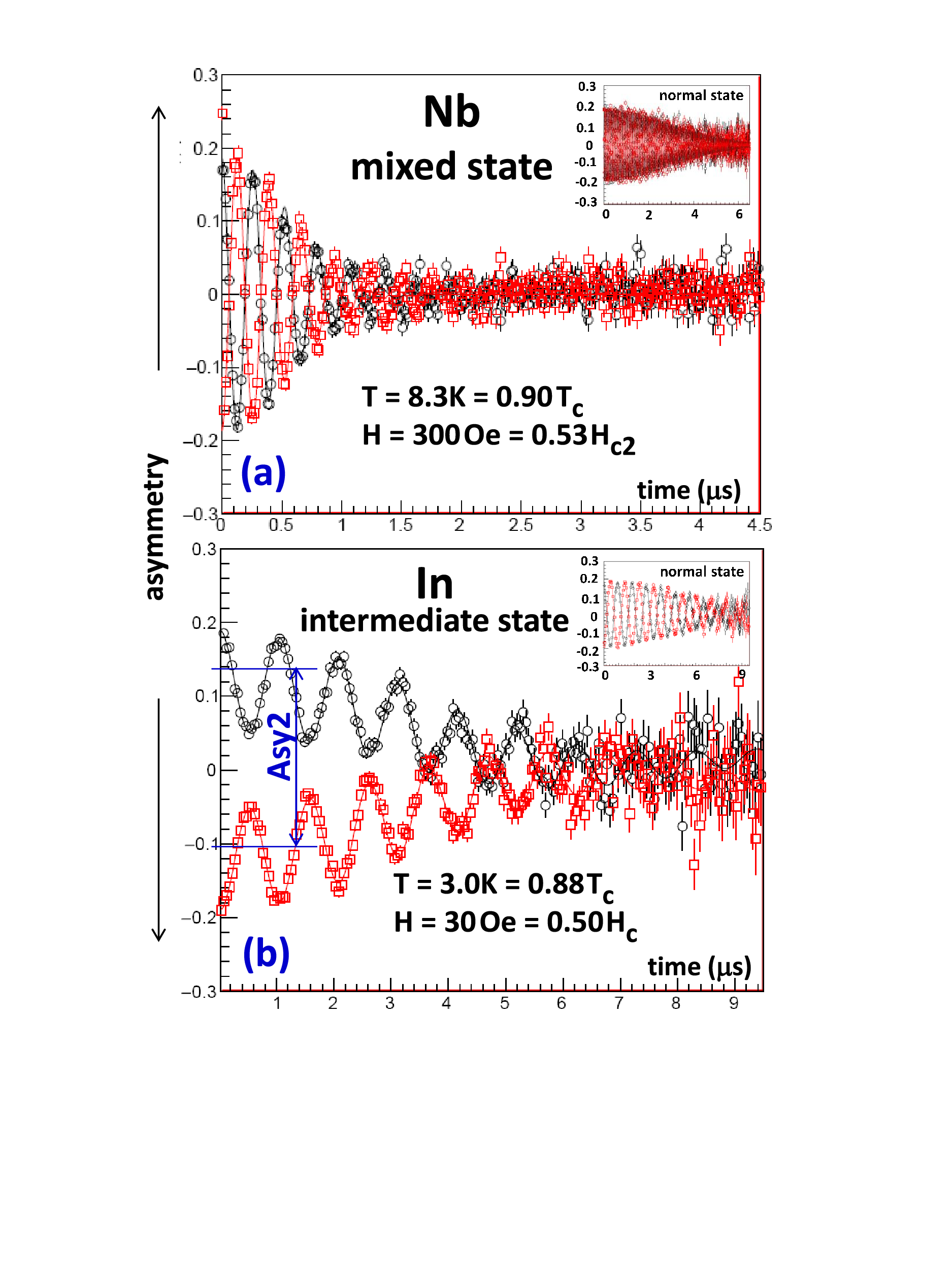,width=5.5 cm}
\caption{\label{fig:epsart} $\mu$SR spectra for (a) single crystal type-II Nb
and (b) single crystal type-I In at the same reduced temperature and field. The inserts show the spectra for the same temperature in the NS. The black (red) dots present the spectra recorded 
along (opposite to) the initial direction of muon spin. $Asy2$ is the asymmetry caused by muons stopping in domains with $B=0$.}
\end{figure}

We see that apart from a much larger damping rate (the damping rate for Nb in the MS normalized relative to that in the NS is greater than the normalized damping rate for In in the IS by a factor of 5), indicating for a strong field inhomogeneity, the spectra of the MS differ from those of the IS by absence of the asymmetry $Asy2$ (see Fig.\,4b), caused by non-precessing muons stopped in S domains with the Meissner ($B=0$)
phase  \cite{Egorov}. Unlike the IS, $B\neq0$ throughout the sample in the MS \cite{Abrikosov, Brandt2003}. Therefore, all muons implanted in such samples precess, resulting in disappearance of $Asy2$.  The absence of  $Asy2$ in $\mu$SR spectra of our Nb sample confirms that it is in the MS but not in the intermediate-mixed state \cite{Forgan, Khasanov}.

\begin{figure}
\epsfig{file=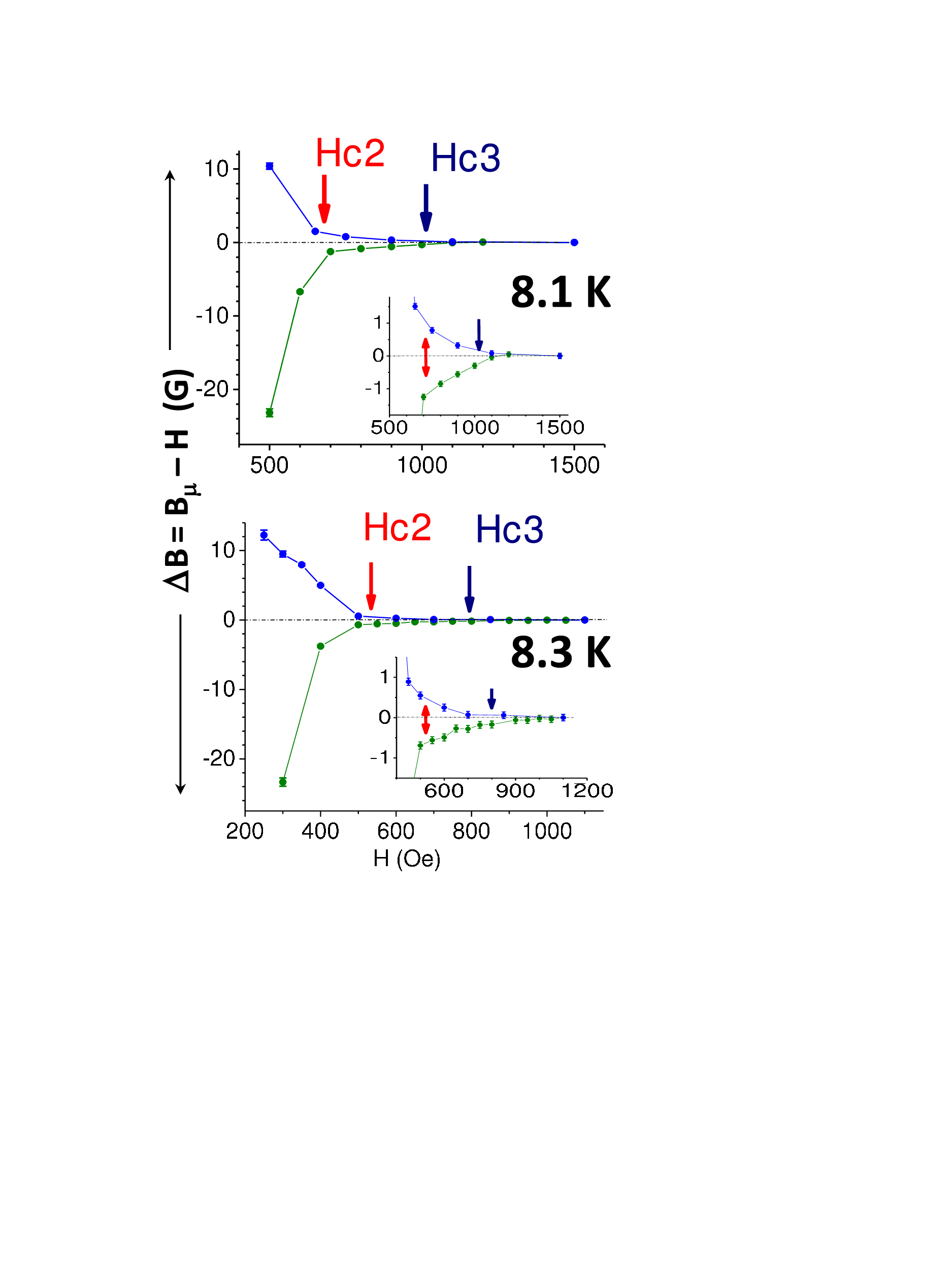,width=4.5 cm}
\caption{\label{fig:epsart}Difference between the $\mu$SR measured $B_\mu$ and the applied
field $H$ vs $H$ at indicated temperatures. 
Green (blue) circles are experimental points obtained at ascending (descending) field. The red and navy arrows indicate 
$H_{c2}$ and $H_{c3}$ obtained from magnetization measurements.}
\end{figure}

Data for the most probable field $B_\mu$ extracted from the $\mu$SR spectra \cite{Sonier} are shown in Fig.\,5 in terms of $\triangle B=B_\mu-H$ vs $H$ on two scales. As seen, $(\partial B_\mu/\partial H)_T$ abruptly changes at $H_{c2}$. At higher field $\triangle B$ decreases vanishing near $H_{c3}$. $H_{c2}$ and $H_{c3}$ were inferred from the $M(H)$ data for the Nb-SC sample. The $\mu$SR data are consistent with those on magnetization apart from a greater hysteresis under descending field, probably caused by a stronger pinning in the Nb-SC2 sample. The $\mu$SR results confirm the presence of the S phase\textit{ in the sample bulk} above $H_{c2}$ \textit{in perpendicular field}, hence supporting the filament scenario.

Images of the magnetic field pattern near the surface of the Nb-F sample were taken using a scanning Hall-probe microscope \cite{Simon}. This was our most challenging experiment due to the low field contrast and the limited microscope resolution. To maximize the signal-to-noise ratio, the images were taken at the lowest possible fields, i.e. at a temperature (9.20 K) very close to $T_c$. 

Typical images are shown in Fig.\,6, where the colors reflect the relative magnitude of the induction, the brightest color corresponds to the strongest $B$. We see that while vortices are clearly distinguishable in a weak field, they become practically unresolvable as $H_{c2}$ is approached. However, a field contrast exceeding the noise level remains below and above $H_{c2}$. To quantify this observation we calculated $B_{rms}=\sqrt{<(B-<B>)^2>}$, where $<...>$ represents a statistical average over the scanned area (7.6 $\times$ 7.6 $\mu$m$^2$). The graphs for  $B_{rms}(H)$ are shown in Fig.\,6, where $H_{c2}$ is inferred from $M(H)$. $B_{rms}\neq 0$ above $H_{c2}$ and it decreases with increasing $H$. This agrees with the data on $M(H)$ and $B_\mu(H)$, confirming that the tiny contrast in the SHPM images above $H_{c2}$ is a real feature consistent  with the filament interpretation.

We conclude that (a) all obtained results are in line with each other; (b) the $M(H)$ and $\mu$SR data reveal the presence of the S phase above $H_{c2}$ in perpendicular field at the same field range as in parallel field; (c) the resistivity data indicate that the S phase forms filaments parallel to the applied field; (d) the filament interpretation is also consistent with the SHPM images.

Now we turn to the question of what is happening near $H_{c2}$. First we note that contrary to the IS, where $M(H)$ \cite{MM} and $B_\mu(H)$ \cite{Egorov} exhibit strong supercooling at the critical field, in the MS, as seen from Figs.\,2 and 5, both $M(H)$ and $B_\mu(H)$ are continuous functions exhibiting discontinuous change in 
$(\partial M/\partial H)_T$ and $(\partial B_\mu/\partial H)_T$ at $H_{c2}$. $M$ and $B$ are the first derivatives of the thermodynamic potentials $\tilde{F}_M(T, V, H)$ and $\tilde{F}(T, V, H^i)$, respectively ($H^i$ is the field strength inside the sample) \cite{Landafshitz_II}. Therefore our results meet the classical definition of second order phase transition \cite{Ehrenfest}, thus confirming the standard interpretation of the transition at $H_{c2}$.

Next, since $M(H)$, $B_\mu(T)$ \cite{Khasanov} and the heat capacity $C(T)$ \cite{Serin} are smooth functions in the MS,  the equilibrium structure near $H_{c2}$ hardly differs from a periodic lattice of vortices, well verified at low $\bar{B}$ \cite{Essmann}. Therefore the filament structure should also be periodic \cite{LL}.
\begin{figure}
\epsfig{file=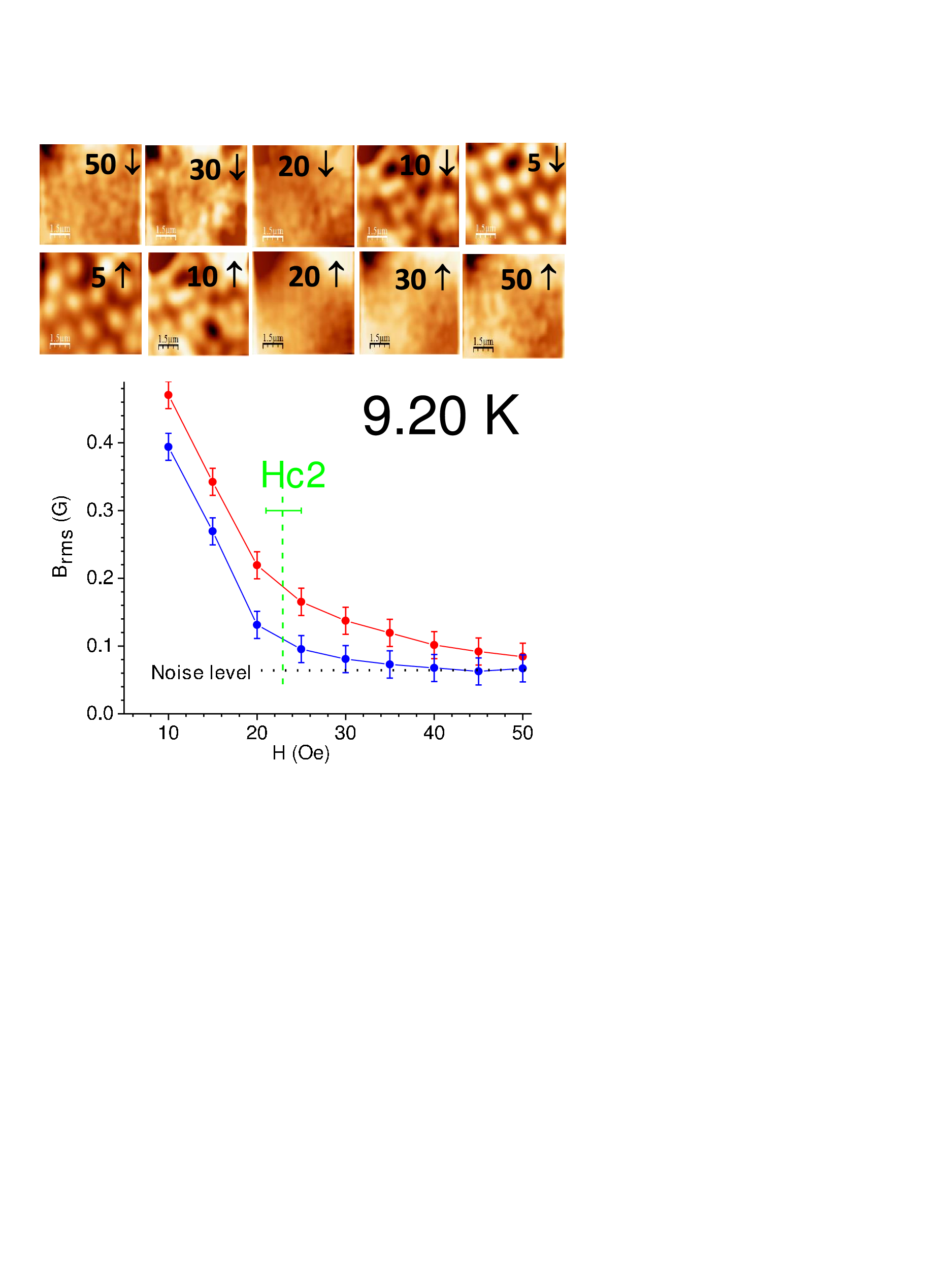,width=5.3 cm}
\caption{\label{fig:epsart}Typical SHPM images of the Nb-F sample, with 
numbers indicating the applied field in Oe. Arrows up (down) indicate images taken at increasing (decreasing) field. The graph presents $B_{rms}(H)$  obtained from the SHPM data. Red (blue) points represent $B_{rms}$ at increasing (decreasing) field. The dashed line designates $H_{c2}$ inferred from $M(H)$ data. 
 }
\end{figure}

Due to hexagonal symmetry of the vortex lattice, a ``landscape" of $B$ has ``peaks" (vortex cores) with maximum $B=H^i$, ``troughs" with minimum $B$ and ``saddle points" between the nearest peaks. Currents form loops about the peaks. The current per unit length of the vortex $g(\varphi, r)$, being a function of the azimuthal ($\varphi$) and radial ($r$) coordinates (see Fig.\,7), is determined by the local gradient of the induction $\partial B/\partial r$ \cite{Landafshitz_II}.  The latter is minimal in the saddle points, thus making these points weak spots in the loops. At $H_{c2}$ the current in the loops ceases. This happens when the angular momentum of electrons in Cooper pairs (or ``superconducting electrons") decreases down to its minimum value, i.e. a quantum of angular momentum $m^*v_r\,r=\hbar$, where $m^*$ is the effective mass of these electrons and $v_r$ is their speed at radius $r$. This Bohr's condition  yields  (see Appendix) the minimum difference $\delta B_{min}$ between the peaks and the saddle points. In CGS units

\begin{eqnarray}
\delta B_{min}=\hbar\frac{4\pi n_s e}{c\,
m^*}\ln\frac{R_\textsf{S}}{R_c}=\frac{\Phi_0}{\pi\lambda_L^2}\ln\frac{R_\textsf{S}}{R_c},
\end{eqnarray}
where $n_s$ and $e$ are number density and charge of the superconducting electrons, $\lambda_L$ is the London penetration depth, $\Phi_0$ is the flux quantum, $R_c$ is the core radius, and $R_\textsf{S}$ is the radius at the saddle point.
\begin{figure}
\epsfig{file=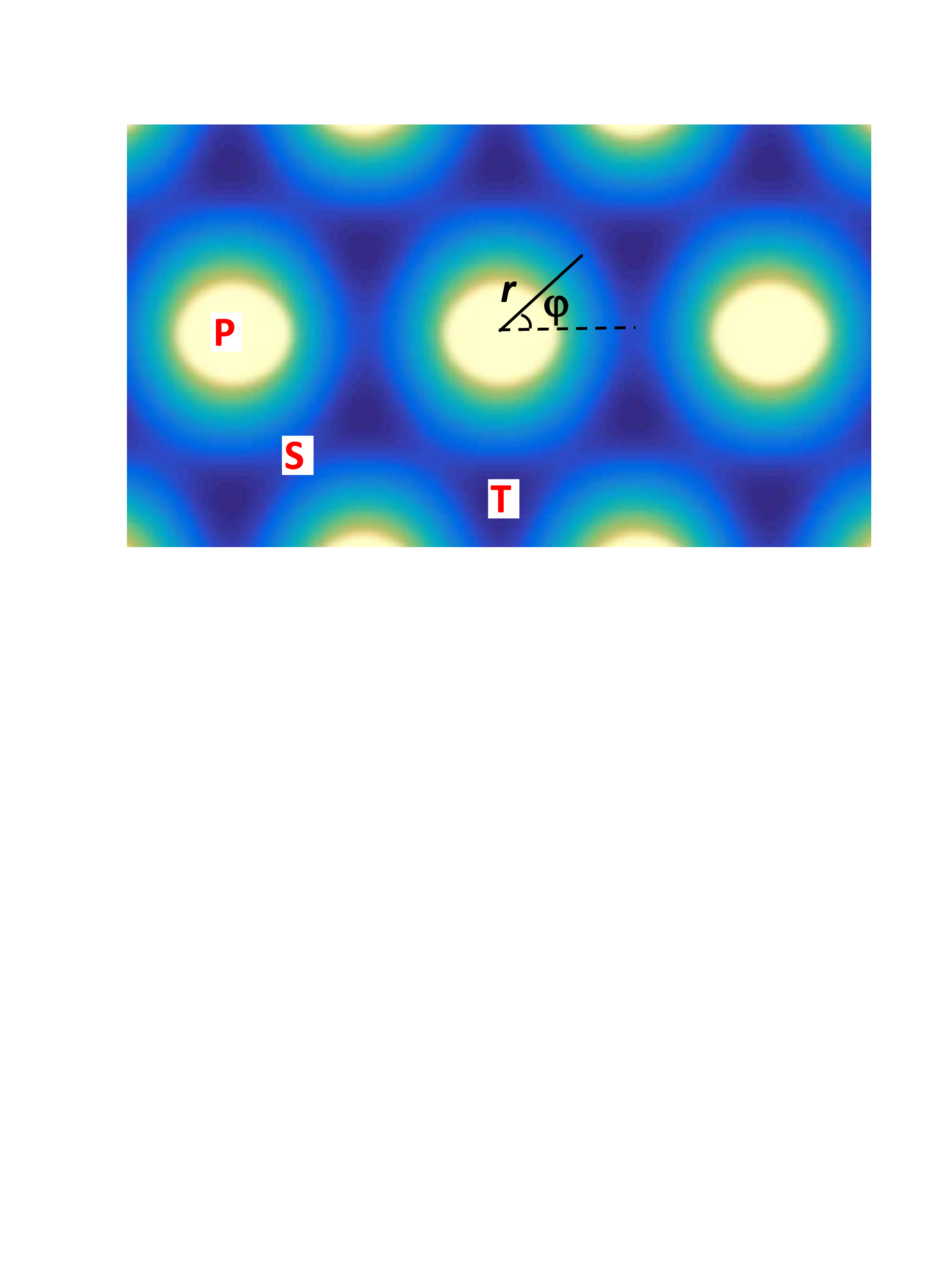,width=5 cm}
\caption{\label{fig:epsart}An induction map of the MS near $H_{c2}$. The field is
perpendicular to the page.  $\textsf{P}$ marks  the ``peaks" with the highest $B=H^i$,
$\textsf{T}$ marks ``troughs" with the lowest $B$, and $\textsf{S}$ marks the ``saddle points"
in between the peaks.}
\vspace{9 mm}
\end{figure}

Hence, consistently with Abrikosov \cite{Abrikosov57}, one can conclude that at $H_{c2}$ the magnetic landscape is not flat. For instance, if $R_\textsf{S}$ differs from $R_c$ by only 0.01\%, $\delta B_{min}$ is already $\sim$1 G. In the troughs $B$ is smaller than in the saddle points, therefore, upon collapse of the vortex current at $H_{c2}$, S phase survives at the troughs where it formes filaments in the out-of-plane (parallel to the field) direction. The amount of S phase just above $H_{c2}$ can be estimated from the difference between the areas of a hexagonal unit cell of the lattice and a circle inscribed in it, which yields about 10\% of the sample volume. Currents driven by the field gradient in the troughs now circulate in the filaments. It is important that right above $H_{c2}$ the filaments keep hexagonal
symmetry of the vortex lattice below $H_{c2}$, thus removing the question about impossibility of the second order phase transition at $H_{c2}$.

Under increasing $H$ the filaments disappear one by one as it happens with S laminae in the IS \cite{IS}. This implies that the filament density continuously decreases down to zero at $H_{c3}$. This is consistent with the data on $M$ (Fig.\,2) and $B_\mu$ (Fig.\,5).

A final point to be addressed is the nucleation of superconductivity under decreasing field. One can expect that the first stable nuclei are small droplets in the sample bulk. Then the field near the droplets is perturbed, making zones of depleted field near the droplet poles. Therefore the next nuclei will preferably appear in these zones, thus creating filaments parallel to the field. In such case the transition at $H_{c3}$ is continuous, in consistent with experiments.

To summarize, results of reported magnetization, electrical transport, $\mu$SR and SHPM measurements performed on Nb samples with different $\kappa$ indicate that superconductivity in type-II materials nucleates at $H_{c3}$ regardless of the orientation of the applied field. Between $H_{c2}$ and $H_{c3}$ superconducting phase exists in the sample bulk, most probably in form of filaments parallel to the applied field.  
Under increasing field above $H_{c2}$ the filament number density decreases vanishing at $H_{c3}$.

The suggested interpretation of properties of type-II superconductors at high field is based on experimental results obtained for two low-$\kappa$ (0.8 and 1.3) superconductors. Therefore it is interesting to verify these observations with materials of higher $\kappa$. Single crystal A15 compounds and high-$T_c$ materials at sufficiently close to $T_c$ temperatures (where pinning is minimal) can be appropriate for such a verification.

\vspace{5 mm}

\maketitle Acknowledgments \vspace{1 mm}

We are grateful to Oscar Bernal and Andrew MacFarlane for the crucial help in organizing the project and to Pabitra Biswas for help in conducting $\mu$SR measurements. This work was supported in part by the National Science Foundation (Grant No. DMR 0904157), by the Research Foundation -- Flanders (FWO, Belgium) and by the Flemish Concerted Research Action (BOF KU Leuven, GOA/14/007) research program. S.J.B. acknowledges support from EPSRC in the UK under grant number EP/J010626/1 and the NanoSC COST Action MP-1201. A.-M. Valente-Feliciano is supported by the U.S. Department of Energy, Office of Science, Office of Nuclear Physics under contract DE-AC05-06OR23177. V.K. acknowledges support from the sabbatical fund of the Tulsa Community College.

\vspace{4 mm}
\textbf{Appendix: Formula (1)}
\vspace{2 mm}


To calculate $\delta B_{min}$ between the peak and the saddle points at $H_{c2}$ we will use cylindrical coordinates with axis parallel to $\textbf{B}$ with azimuthal $\varphi$ and radial $r$ coordinates shown in Fig.\,7.

Change $dB$ in the normal (radial) direction over a radial interval $dr$ (see Fig.\,8) occurs due to the current $dI= ldg$ running in azimuthal direction in the cylindrical layer of radius $r$ and thickness $dr$, where $l$ is the length of the cylinder (length of the vortex) and $g$ is the current per unite length of the cylinder. In CGS units $dB$ and $dg$ are linked as  \cite{Landafshitz_II}
\begin{eqnarray}
dB=\frac{4\pi}{c}dg,
\end{eqnarray}
where $c$ is speed of light.
\begin{figure}
\epsfig{file=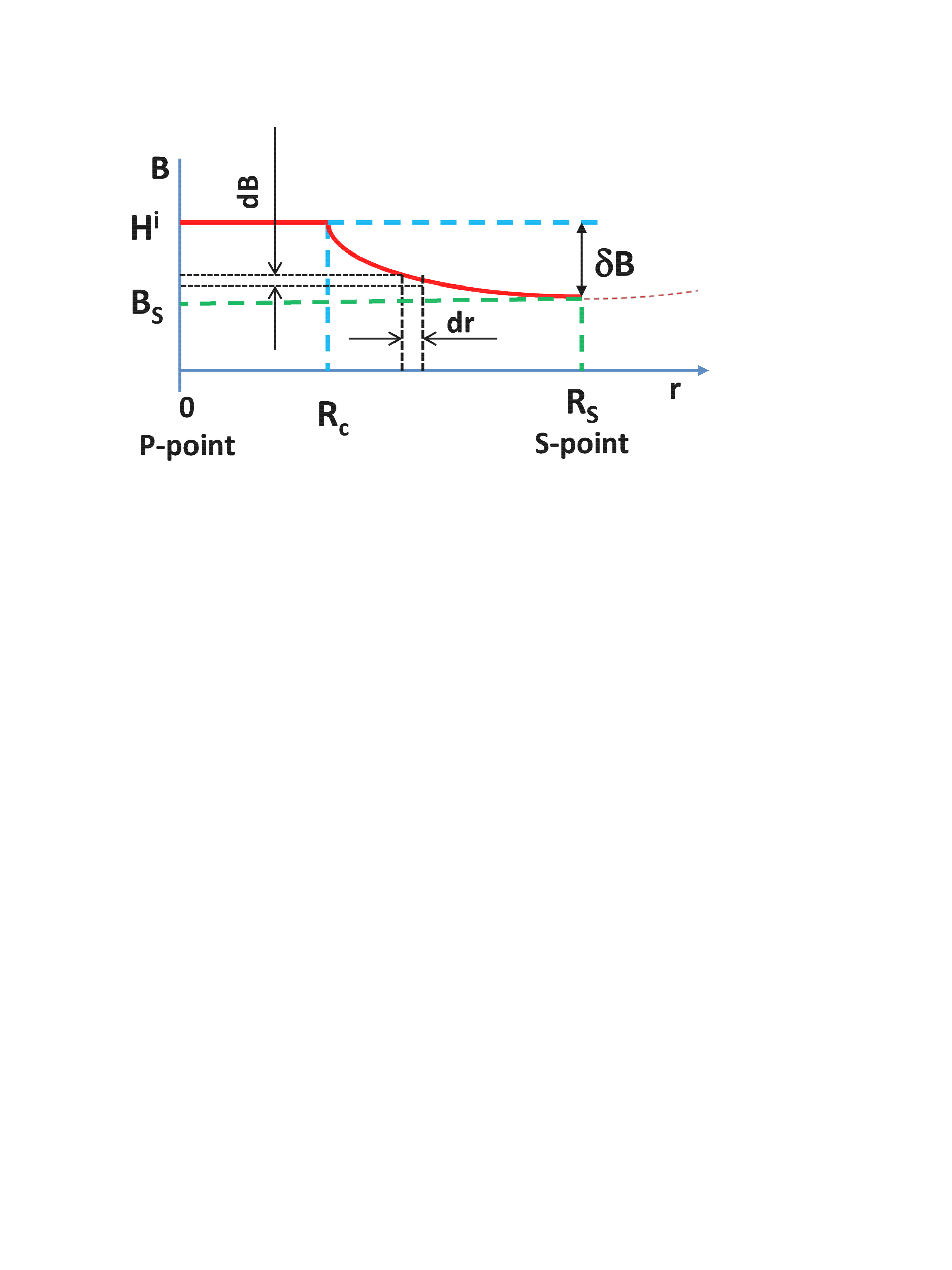,width=5 cm}
\caption{\label{fig:epsart}Profile of induction in a unit cell with a single vortex along
\textsf{P}-\textsf{S} line (see Fig.\,7 of the main material).}
\end{figure}

Therefore,
\begin{eqnarray}
dI=ldg=l\frac{c}{4\pi}dB=(n_s e v_r )ldr,
\end{eqnarray}
where $n_s$, $e$ and $v_r$ is number density, charge and speed of superconducting electrons (electrons paired in Cooper pairs) in the layer, respectively, and $n_s e v_r$ is density of the current running through the cross sectional area $ldr$.

Therefore,
\begin{eqnarray}
n_s e v_r r dr=\frac{c\, r dB}{4\pi}.
\end{eqnarray}

At $H_{c2}$ the Bohr condition for the minimal angular momentum of the superconducting electron is
\begin{eqnarray}
L_{min}=m^*r v_r=\hbar.
\end{eqnarray}

Therefore,
\begin{eqnarray}
dB=\frac{4\pi n_s e (v_r r)}{c\, r}dr|_{at\, H_{c2}}=\hbar \frac{4 \pi n_s e}{c\, r\, m^*}dr.
\end{eqnarray}

Integrating the last expression over the radial interval from the radius of the core $R_c$ to the radius of the saddle point $R_\textsf{S}$ one obtains formula (1)
\begin{multline}
\delta B_{min}=\hbar\frac{4\pi n_s e}{c\,m^*}\int_{R_c} ^{R_\textsf{S}}
\frac{dr}{r}=\hbar\frac{4\pi n_s e}{cm^*}\ln\frac{R_\textsf{S}}{R_c} =\\=\frac{\Phi_0}{\pi\lambda_L^2 }\ln\frac{R_\textsf{S}}{R_c},
\end{multline}
where $\Phi_0$ and $\lambda_L$ are the superconducting flux quantum and the London penetration depth, respectively.

\end{document}